\documentclass[aps,prd,preprint,groupedaddress,showpacs,showkeys,elide]{revtex4}
\usepackage{bm,graphicx,amssymb}
\newcommand{\crist}[3]{\ensuremath{\left\{^{\,\,#1}_{#2#3}\right\}}}

\newcommand{\diag}{\mbox{diag}}
\begin{document}
\title{\textit{Ab initio} derivation of the quantum Dirac's equation by conformal differential geometry:  the ``Affine Quantum Mechanics''}
\author{Enrico Santamato}
\email{enrico.santamato@na.infn.it}
\affiliation{Dipartimento di Scienze Fisiche, Universit\`{a} di
Napoli ``Federico II'', Compl.\ Univ.\ di Monte S. Angelo, 80126
Napoli, Italy}
\affiliation{Consorzio Nazionale Interuniversitario per le Scienze
Fisiche della Materia, Napoli}
\author{Francesco De Martini}
\email{francesco.demartini@uniroma1.it}
\affiliation{Dipartimento di Fisica dell'Universit\`{a} ``La
Sapienza'' and Consorzio Nazionale Interuniversitario per le Scienze
Fisiche della Materia, Roma 00185, Italy}
\affiliation{Accademia Nazionale dei Lincei, via della Lungara 10,
Roma 00165, Italy}
\begin{abstract}
A rigorous \textit{ab initio} derivation of the (square of) Dirac's equation for a single particle with spin is presented. The general Hamilton-Jacobi equation for the particle expressed in terms of a background Weyl's conformal geometry is found to be linearized, exactly and in closed form, by an \textit{ansatz} solution that can be straightforwardly interpreted as the ``quantum wave function'' $\psi_4$ of the 4-spinor Dirac's equation. In particular, all quantum features of the model arise from a subtle interplay between the conformal curvature of the configuration space acting as a potential and Weyl's ``pre-potential'', closely related to $\psi_4$, which acts on the particle trajectory. The theory, carried out here by assuming a Minkowsky metric, can be easily extended to arbitrary space-time Riemann metric, e.g. the one adopted in the context of General Relativity. This novel theoretical scenario, referred to as ``Affine Quantum Mechanics'', appears to be of general application and is expected to open a promising perspective in the modern endeavor aimed at the unification of the natural forces with gravitation.
\end{abstract}

\maketitle


\section{Introduction}
The particle spin theory is one of the cornerstones of quantum mechanics. Consequently, being the spin a peculiar feature of the quantum world, any attempt to find a classical system behaving as a spinning quantum particle is generally considered hopeless. In this work, we show that a wave equation for quantum spin (and, in particular, the square of Dirac's spin 1/2 equation) may be derived from the mechanics of the relativistic top in a curved configuration space.\\ Our approach is based on the theory proposed some time ago by one of us to derive the Schr\"{o}dinger and the Klein-Gordon equations from mechanics in curved spaces~\cite{santamato84,*santamato85}. The main physical assumption of this theory, we shall refer to henceforth as ``Affine Quantum Mechanics'' (AQM), is that the origin of quantum effects is a feedback between geometry and dynamics~\footnote{According to the Felix Klein Erlangen program, the affine geometry deals with intrinsic geometric properties that remain unchanged under affine transformations (affinities). These preserve collinearity transformations e.g. sending parallels into parallels, and ratios of distances along parallel lines. The Weyl geometry is considered a kind of affine geometry preserving angles. Cfr: H. Coxeter, Introduction to Geometry (Wiley, New York 1969).}. More precisely, in this work we ascribe the quantum effects to the presence of a not trivial parallel transport law in the particle configuration space bearing a nonzero curvature. The metric has no role here and can be arbitrarily prescribed. In fact, at the very fundamental level, the space curvature is originated by the affine connections and not by the metric tensor. The AQM assumes that the actual affine connections (and hence the scalar curvature) of the particle configuration space are affected by the dynamics of the particle itself and that, in turn, the space scalar curvature acts on the particle as a potential. Thus, the particle motion and the space affine connections must be determined consistently. The overall physical picture is analogous to the situation prevailing in general relativity: geometry is not prescribed; rather it is determined by the physical reality. In turn, geometry acts as a ``guidance field'' for matter. The idea of a ``guidance field'' to explain quantum phenomena dates back to De Broglie. The AQM identifies the origin of the already mysterious De Broglie field with the curvature of space-time and casts its effects on a firm and plausible theoretical frame. However, unlike in general relativity, the space curvature is ascribed by AQM to the affine connections rather than to the metric of the geometry. In this way, gravitational and quantum phenomena share a common geometrical origin, but are based on \textit{independent} geometrical objects: the metric tensor for the former and the affine connections for the latter. As in general relativity, the geometric approach forces to describe matter as a fluid or as a bundle of elementary trajectories rather than as a single point particle moving along one trajectory. In this respect, AQM is somewhat related to the hydrodynamic approach to quantum mechanics first proposed by Madelung~\cite{madelung26} and then developed by Bohm~\cite{bohm83a, *bohm83b}. In the Madelung-Bohm approach the particle trajectories are deterministically governed by classical mechanics and quantum effects are due to a ``quantum potential'' of quite mysterious origin, whose gradient acts as a newtonian force on the particle. According to the AQM the active potential originates from geometry, as does gravitation, and arises form the space curvature due to the presence of the non trivial affine connections of the Weyl conformal geometry~\cite{weyl18,*weyl}.
\section{The relativistic top}
We start considering the simplest model for the relativistic spinning particle, namely the top described by six Euler angles, as made, for example, by Frenkel~\cite{frenkel26}, Thomas~\cite{thomas26} and in the classic work by Hanson and Regge~\cite{hanson74}. Subsequent important works on relativistic spinning particles can be found in many textbook~\cite{souriau,corben,sudarshan}. Here, we simply imagine that the particle follows a path $x^\mu=x^\mu(\sigma)$ in space-time, where $\sigma$ is an arbitrary parameter along the path, and that it carries along with itself a moving fourleg $e^\mu_a=e^\mu_a(\sigma)$ ($\mu,a=0,\dots,3$). The fourleg vectors $e^\mu_a$ are normalized according to $g_{\mu\nu}e^\mu_ae^\nu_b=g_{ab}$ where $g_{\mu\nu}=g_{ab}=\diag(-1,1,1,1)$ is the Minkowski metric tensor. Setting $\Lambda(\sigma)=\{e^\mu_a(\sigma)\}$ and $G=\{g_{\mu\nu}\}$, the normalization relations can be cast in the matrix form as $\Lambda^T G\Lambda=G$, showing that the $4{\times}4$ matrix $\Lambda(\sigma)\in SO(3,1)$ is a proper Lorentz matrix. The derivative of $e^\mu_a(\sigma)$ with respect to $\sigma$ can be written as $de^\mu_a/d\sigma=\omega^\mu_\nu e^\nu_a$. The contravariant tensor $\omega^{\mu\nu}=\omega^\mu_\sigma g^{\sigma\nu}$ is skewsymmetric and can be considered as the ``angular velocity'' of the top in space-time. The free Lagrangian of this minimal relativistic top is
\begin{eqnarray}\label{eq:L0}
  L_0 &=& mc\sqrt{-g_{\mu\nu}\frac{dx^\mu}{d\sigma} \frac{dx^\nu}{d\sigma} -
             a^2 g_{\mu\nu}g^{ab}\frac{de^\mu_a}{d\sigma}\frac{de^\nu_b}{d\sigma}}=
             \nonumber \\
        &=& mc\sqrt{-g_{\mu\nu}\frac{dx^\mu}{d\sigma} \frac{dx^\nu}{d\sigma} -
             a^2 \omega_{\mu\nu} \omega^{\mu\nu}},
\end{eqnarray}
where $m$ is the particle mass, $c$ is the speed of light, and $a$ is a constant having the dimension of a length. For a quantum particle of mass $m$ we expect $a$ to be of the order of the particle Compton wavelength. The square root in Eq.~(\ref{eq:L0}) ensures that $L_0$ is parameter invariant. In the presence of an external electromagnetic field, the total Lagrangian becomes $L=L_0+L_{em}$, where the electromagnetic interaction Lagrangian is taken as
\begin{equation}\label{eq:Lem}
  L_{em} = \frac{e}{c}A_\mu \frac{dx^\mu}{d\sigma} +
        \frac{\kappa e}{4c}\;a^2 F_{\mu\nu}\omega^{\mu\nu}
\end{equation}
where $e$ is the particle charge and $F_{\mu\nu}$ is given by $F_{\mu\nu}=\partial A_\nu/\partial x^\mu-\partial A_\mu/\partial x^\nu$ with four-potential $A_\mu$ given by $A_\mu=(-\phi,\bm A)$, $\phi$, $\bm A$ being the scalar and vector electromagnetic potentials, respectively. Finally, $\kappa$ is a numeric constant that will be identified as the particle gyromagnetic ratio. The fourleg components $e^\mu_a$ (and the $SO(3,1)$ group) are parametrized by six ``Euler angles'' $\theta^\alpha$ ($\alpha=1,\dots,6$), so that the configuration space spanned by the space-time coordinates and the Euler angles is ten dimensional. When $\omega^{\mu\nu}$ is written in terms of the angles $\theta^\alpha$ and their derivatives, the free-particle Lagrangian $L_0$ assumes the standard form
\begin{equation}\label{eq:L0geo}
  L_0=mc\frac{ds}{d\sigma}=
      mc\sqrt{-g_{ij}\frac{dq^i}{d\sigma}\frac{dq^j}{d\sigma}},
\end{equation}
where $q^i=\{x^\mu,\theta^\alpha\}$ $(i=0,\dots,9)$ are the ten coordinates spanning the dynamical configuration space of the top~\footnote{The configuration space of the top described by the Lagrangian $L$ is the principal fiber bundle whose base is the Minkowski space-time ${\cal M}_4$ and whose fiber is $SO(3,1)$, conceived as a proper Lorentz frame manifold. The dynamical invariance group is the whole Poincar\'{e} group of the inhomogeneous Lorentz transformations.}. Similarly, the electromagnetic interaction Lagrangian $L_{em}$ assumes the standard form $L_{em}=(e/c)A_idq^i/d\sigma$, where $A_i=(A_\mu,A_\alpha)$ is a ten dimensional covariant vector. The last six entries $A_\alpha$ of $A_i$ are linear combinations of the components of the magnetic and electric
fields $\bm H(x)$ and $\bm E(x)$, respectively~\footnote{More specifically, $A_\alpha=\xi^a_\alpha(\theta)A_a(x)$
$(\alpha,a=1,\dots,6)$, where $A_a(x)=-\frac{\kappa}{2}\{\bm H(x),\bm E(x)\}$ and $\xi^a_\alpha(\theta)$ are the Killing vectors of the Lorentz group $SO(3,1)$.}. The quantities $g_{ij}$ in Eq.~(\ref{eq:L0geo}) define the distance $ds=\sqrt{-g_{ij}dq^idq^j}$ in the top configuration space, which is so converted into the 10-D Riemann metric space $V_{10}={\cal M}_4{\times}SO(3,1)$. The extremal curves of $L_0$ are the geodetics of this space. The metric tensor $g_{ij}$ in $V_{10}$ has the diagonal block form
  $g_{ij}=\pmatrix{
                g_{\mu\nu} & 0 \cr
                0 & g_{\alpha\beta}
                }$,
where $g_{\mu\nu}$ is the Minkowski metric and $g_{\alpha\beta}$ is the metric of the parameter space of Lorentz group with signature $(+,+,+,-,-,-)$ and we assumed the Euler angles $\theta^\alpha$ ordered so that the first three angles $\theta^\alpha$ for $\alpha=1,2,3$ are associated with space rotations, and the last three angles $\theta^\alpha$ for $\alpha=4,5,6$ to Lorentz boosts. The classical mechanics induced by the extremals of the Lagrangian $L$ on the space $V_{10}$ is well known~\cite{souriau,corben,sudarshan}. Here it is enough noticing that neither the time-like vector $e^\mu_0$ of the moving fourleg is identified by the particle four-velocity $u^\mu=dx^\mu/d\tau$, $d\tau=\sqrt{-g_{\mu\nu}dx^\mu dx^\nu}$ being the proper time, nor Weysenhoff's kinematical constraints $\omega_{\mu\nu}u^\nu=0$ are imposed, in general. As a consequence, the so called ``center-of-mass'' space-time trajectory $x^\mu(\tau)$ and the so called ``center-of-energy'' space-time trajectory $y^\mu(\tau)$ of our top (obtained from $dy^\mu/d\tau=e^\mu_0$) are different~\cite[see Ref.~][chap.~20]{sudarshan}. The main advantage of using a top described by six Euler angles is that the usual methods of analytical mechanics can be applied without worrying about kinematical constraints; but the usual picture of spin as the coadjoint action of the little Poincar\'{e} group on the particle momentum space~\cite[see Ref.~][Chap.~3, sec.~13]{souriau} is generally lost.
\section{The conformal relativistic top}
The main result of the present work is to show that the square of Dirac's equation for the quantum spin 1/2 particle can be obtained by a simple change of Lagrangian so to provide Weyl's conformal invariance to the particle dynamics without introducing any concept extraneous to the classical world. In other words, to describe the quantum spinning particle, we propose to use in place of the Lagrangian $L=L_0+L_{em}$ a new Lagrangian $\bar L$ which is invariant under the conformal change $g_{ij}\rightarrow\rho(q)g_{ij}$ of the configuration space metric. As suggested elsewhere for spinless particles~\cite{santamato84,*santamato85}, we introduce conformal invariance by assuming that the configuration space of the top is a Weyl space with metric $g_{ij}$ and integrable Weyl's connections $\Gamma^i_{jk}$ given by
\begin{equation}\label{eq:GammaWeyl}
  \Gamma^i_{jk}=-\crist{i}{j}{k}+
      \delta^i_j\phi_k+\delta^i_k\phi_j+g_{jk}\phi^i,
\end{equation}
where $\crist{i}{j}{k}$ are the Cristoffel symbols out of the metric $g_{ij}$, $\phi^i=g^{il}\phi_l$, and $\phi_i$ is the Weyl potential that we assume to be integrable, viz. $\phi_i=\chi^{-1}\partial\chi/\partial q^i$. As Lagrangian we take
\begin{equation}\label{eq:Lbar}
  \bar L =
      \xi\hbar\sqrt{-R_W g_{ij}\frac{dq^i}{d\sigma}\frac{dq^j}{d\sigma}} + L_{em},
\end{equation}
where $\xi$ is a numeric constant, $L_{em}$ is given by Eq.~(\ref{eq:Lem}) and $R_W$ is the Weyl scalar curvature calculated from the
connections (\ref{eq:GammaWeyl}), viz.
\begin{eqnarray}\label{eq:RW}
  R_W &=& R + 2(n-1)\nabla_k\phi^k-(n-1)\phi_k\phi^k = \nonumber\\
      &=& R + 2(n-1)\frac{\nabla_k\nabla^k\chi}{\chi} -
       n(n-1)\frac{\nabla_k\chi\nabla^k\chi}{\chi^2},
\end{eqnarray}
where $\nabla_i$ denote the covariant derivatives built out from the Cristoffel symbols {\crist{i}{j}{k}}. The Lagrangian (\ref{eq:Lbar}) is manifestly invariant under the conformal changes of the metric and Weyl potential $g_{ij}\rightarrow\rho g_{ij}$ and $\phi_i\rightarrow\phi_i-\rho^{-1}\partial\rho/\partial q^i$, respectively, provided the fields $A_i$ are Weyl invariant. The dynamic theory derived from the Lagrangian $\bar L$ applying the extremal action principle is conformally invariant too. Moreover, it is worth noting that the Lagrangian $\bar L$ is massless, because the particle mass $m$ was replaced by Weyl's curvature according to $mc\rightarrow \xi\hbar \sqrt{R_W}$. We will call the top described by the  Lagrangian $\bar L$ the conformal relativistic top. The Weyl curvature field $R_W(q)$ in Eq.~(\ref{eq:RW}) acts as a scalar potential on the conformal top and, because it depends on $\chi$ and its derivatives, the field $\chi$ acts on the conformal top as a sort of pre-potential. The paths followed by the conformal top in the configuration space $V_{10}={\cal M}_4{\times}SO(3,1)$ are the extremal curves of the action integral $\int \bar L\:d\sigma$. Of particular importance are the bundles of extremals belonging to a family of equidistant hypersurfaces $S=\mathrm{const.}$ in the configuration space. These bundles are obtained from the solutions of the Hamilton-Jacobi equation associated with $\bar L$
\begin{eqnarray}\label{eq:HJbar}
     g^{ij}
      \left(\frac{\partial S}{\partial q^i}-\frac{e}{c}A_i\right)
      \left(\frac{\partial S}{\partial q^j}-\frac{e}{c}A_j\right)= \nonumber\\
    = g^{ij}\left(D_i S-\frac{e}{c}A_i\right)
      \left(D_j S -\frac{e}{c}A_j\right) = -\hbar^2\xi^2 R_W
\end{eqnarray}
by integrating the differential equations
\begin{equation}\label{eq:dqds}
   \frac{dq^i}{ds}= \frac{g^{ij}(\frac{\partial S}{\partial q^j}-\frac{e}{c}A_j)}
       {[g^{mn}(\frac{\partial S}{\partial q^m}-\frac{e}{c}A_m)
       (\frac{\partial S}{\partial q^n}-\frac{e}{c}A_m)]^{1/2}}.
\end{equation}
Moreover, we assume that the action function $S$ obeys the auxiliary divergence condition
\begin{equation}\label{eq:div}
  D_k\left(D^k S-\frac{e}{c}A_k\right) = 0.
\end{equation}
We may think this condition as stating that the trajectories in the bundle do not intersect in the considered region of the configuration space. In Eqs.~(\ref{eq:HJbar}) and (\ref{eq:div}) $D_i$ denote the Weyl co-covariant derivatives with respect of the coordinate $q^i$~\footnote{The action of $D_i$ over a tensor field $F$ of Weyl type $w(F)$ is given by $D_i F= \nabla^{(\Gamma)}_i F -2w(F)\phi_iF$, where $\phi_i$ is the Weyl potential and $\nabla^{(\Gamma)}_i$ is the covariant derivative built up from the Weyl connections $\Gamma^i_{jk}$ given by Eq.~(\ref{eq:gamma}). The Weyl type of $D_i F$ is the same as of $F$.}. The use of the co-covariant derivatives makes explicit the coordinate and conformal gauge covariance of Eqs.~(\ref{eq:HJbar}) and (\ref{eq:div}). When written out in full, Eq.~(\ref{eq:div}) states the conservation of the Weyl-invariant current density
\begin{equation}\label{eq:j}
   j^i=\chi^{-(n-2)}\sqrt{g}\:g^{ij}(\partial S/\partial q^j-(e/c)A_j).
\end{equation}
Equations~(\ref{eq:HJbar}) and (\ref{eq:div}) are a set of nonlinear partial differential equations for the unknown functions $S(q)$ and $\chi(q)$, once the metric tensor $g_{ij}(q)$ is given. The nonlinear problem implied by Eqs.~(\ref{eq:HJbar}) and (\ref{eq:div}) looks very hard at first glance. However, bu introducing the complex scalar function $\psi$ of Weyl'e type $w(\psi)=-(n-2)/4$ given by
\begin{equation}\label{eq:psidef}
  \psi(q)=\chi(q)^{-\frac{n-2}{2}}e^{i\frac{S(q)}{\hbar}}
\end{equation}\
and fixing $\xi$ according to
\begin{equation}\label{eq:gamma}
  \xi^2=\frac{n-2}{4(n-1)}=\frac{2}{9},
\end{equation}
where $n=10$ is the dimensionality of the top configuration space, converts Eqs.~(\ref{eq:HJbar}) and (\ref{eq:div}) into the \textit{linear} differential equation
\begin{equation}\label{eq:KG}
  g^{ij}\left(\hat p_i-\frac{e}{c}A_i\right)
         \left(\hat p_j-\frac{e}{c}A_j\right)\psi+
         \hbar^2 \xi^2 R\psi=0.
\end{equation}
where $\hat p_=-i\hbar \nabla_i$. This is a striking result as it demonstrates that the Hamilton-Jacobi equation, applied  to a general dynamical problem can be transformed into a linear eigenvalue equation, the foremost ingredient of the formal structure of quantum mechanics and of the Hilbert space theory. Note that the transition from the Hamilton-Jacobi Eq.~(\ref{eq:HJbar}) to the quantum mechanical Eq.~(\ref{eq:KG}) implies the adoption of a formally trivial albeit conceptually significant  transformation: $p_i = \partial S/\partial x^i \rightarrow (-i\hbar\partial/\partial x^i) \times(i S(x)/\hbar)$, where the two factors are commonly interpreted as momentum operator $(\hat p)$ and (complex) phase, respectively. This transformation precisely represents the transition from Hamilton's classical dynamics to quantum mechanics in our theory. In the absence of the electromagnetic field ($A_i=0$), Eq.~(\ref{eq:KG}) reduces to $\hat L\psi=(-\Delta+\xi^2 R)\psi$, where $\Delta$ is the Laplace-Beltrami operator and $\hat L$ is the conformal Laplacian, also known as the Laplace-de Rham operator associated with the metric $g_{ij}$. The value of $\xi$ given by Eq.~(\ref{eq:gamma}) ensures that Eq.~(\ref{eq:KG}) is conformally invariant. The Laplace-de Rham Eq.~(\ref{eq:KG}) resembles the covariant quantum Klein-Gordon wave-equation in the configuration space with the mass term $m^2c^2$ replaced by the curvature potential term $\hbar^2\xi^2 R(q)$. What it is more surprising is that any explicit reference to the Weyl pre-potential $\chi(q)$ and to the Weyl curvature $R_W$ has been cancelled out from Eq.~(\ref{eq:KG}). In fact, the curvature $R(q)$ and the covariant derivatives $\nabla_i$ in Eq.~(\ref{eq:KG}) are calculated using the Cristoffel symbols derived from the metric $g_{ij}$. Finally, the (Riemann) curvature $R$ of the top configuration space is constant in our case, and it is given by $R=6/a^2$. Moreover, the conserved current density $j^i$ in Eq.~(\ref{eq:j}) can be written in the alternative form
\begin{equation}\label{eq:jalt}
   j^i=|\psi|^2\sqrt{g}\:g^{ij}(\partial S/\partial q^j-(e/c)A_j).
\end{equation}
without any explicit reference to the underlying Weyl's geometry. The current (\ref{eq:jalt}) together with Eq.~(\ref{eq:dqds}) shows that the scalar density $|\psi|^2$ is transported along the particle trajectory in the configuration space, allowing the optional statistical interpretation of the wavefunction $\psi$ according to Born's quantum mechanical rule~\cite{santamato84,*santamato85}. The reduction of Eqs.~(\ref{eq:HJbar}) and (\ref{eq:div}) to the wave-equation (\ref{eq:KG}) is the central result of this work, because it builds a bridge between the quantum and the classical worlds. The quantum wave equation (\ref{eq:KG}) with the $|\psi|^2$ Born prescription is mathematically equivalent to the classical Hamilton-Jacobi Eq.~(\ref{eq:HJbar}) associated with the conformally invariant Lagrangian $\bar L$; Born's rule comes out in a very natural way from the conformally invariant zero divergence current requirement along any Hamiltonian bundle of trajectories in the configuration space. It is also worth noting that Born's rule relays on the particular choice of the conformal gauge made to obtain the Laplace-de Rham Eq.~(\ref{eq:KG}). By changing the gauge, we can make $|\psi|^2\rightarrow |\bar\psi|^2=1$ in which case the Weyl curvature reduces to the Riemann curvature $\bar R$ of the not trivial metric $\bar g_{ij}=\chi^{-2}g_{ij}=|\psi|^{4/(n-2)}g_{ij}$. Our final step is now to show that the wave equation~(\ref{eq:KG}) is able to account for the quantum spin 1/2.\\
\section{Equivalence with Dirac's equation}
We first note that Eq,~(\ref{eq:KG}) is invariant under parity $P$, so we may look for solutions $\psi(q)$ which also are invariant under $P$. These solutions can be cast in the mode expansion form
\begin{eqnarray}\label{eq:psiuv}
  \psi_{uv}(q) =
    D^{(u,v)}(\Lambda^{-1})^\sigma_{\sigma'}\psi^{\sigma'}_{\sigma}(x)&+&
    D^{(v,u)}(\Lambda^{-1})^{\dot\sigma}_{\dot\sigma'}
    \psi^{\dot\sigma'}_{\dot\sigma}(x)\nonumber \\
    &&(u\le v)
\end{eqnarray}
where $D^{(u,v)}(\Lambda)^\sigma_{\sigma'}$ are the $(2u+1){\times}(2v+1)$ matrices representing the Lorentz transformation
$\Lambda(\theta)=\{e^\mu_a(\theta)\}$ in the irreducible representation labeled by the two numbers $u,v$ given by
$2u,2v=0,1,2,\dots$, and the $\psi^{\sigma'}_\sigma(x)$ and $\psi^{\dot\sigma'}_{\dot\sigma}(x)$ are expansion coefficients depending on the space-time coordinates $x^\mu$ alone. The matrices $D^{(u,v)}(\Lambda)$ and $D^{(v,u)}(\Lambda)$ depend on the Euler angles $\theta^\alpha$ only, and provide conjugate representations of
the Lorentz transformations~\footnote{The two matrices are related by $[D^{(u,v)}(\Lambda)]^\dag=[D^{(v,u)}(\Lambda)]^{-1}$.}. As the notation suggests, the invariance of $\psi_{uv}(q)$ under Lorentz transformations implies that $\psi^{\sigma'}_{\sigma}(x)$ and $\psi^{\dot\sigma'}_{\dot\sigma}(x)$ change as undotted and dotted contravariant spinors, respectively~\footnote{The spinors $\psi^{\sigma'}_{\sigma}(x)$ and $\psi^{\dot\sigma'}_{\dot\sigma}(x)$ are invariant with respect to their lower indices, which are related
to the spin component along the top axis.}. In Eq.~(\ref{eq:psiuv}) both dotted and undotted spinors appear, because we are interested in solutions $\psi_{uv}(q)$ of Eq.~(\ref{eq:KG}) which are parity invariant. Indeed, both terms on the right of Eq.~(\ref{eq:psiuv}) obey Eq.~(\ref{eq:KG}) separately, each one providing not parity invariant solutions to Eq.~(\ref{eq:KG}). In the case of spin 1/2, the spinors $\psi^{\sigma'}_{\sigma}(x)$ and
$\psi^{\dot\sigma'}_{\dot\sigma}(x)$ have two components. The use of two-components spinors in place of the four-component Dirac's spinors have been extensively discussed in the literature~\cite{brown58}. In
this paper, however, we will limit to parity invariant solutions of Eq.~(\ref{eq:KG}) described by four-component Dirac's spinors. Insertion of the expansion (\ref{eq:psiuv}) into the wave-equation (\ref{eq:KG}) yields to the following equation for the coefficients $\psi^{\sigma'}_{\sigma}(x)$ and $\psi^{\dot\sigma'}_{\dot\sigma}(x)$
\begin{eqnarray}\label{eq:coeff}
       \left[g^{\mu\nu}\left(\hat p_\mu -\frac{e}{c}A_\mu\right)
         \left(\hat p_\nu -\frac{e}{c}A_\nu\right)\right.
         &+&\left. \hbar^2\xi^2 R\right]\psi(x) +\nonumber \\
         \mbox{} + \Delta_J \psi(x) &=& 0
\end{eqnarray}
where $\hat p_\mu=-i\hbar\partial_\mu$ and $\psi(x)$ denotes either $\psi^{\sigma'}_{\sigma}(x)$ or $\psi^{\dot\sigma'}_{\dot\sigma}(x)$. Finally, $\Delta_J$ is a $(2u+1){\times}(2v+1)$ matrix depending on the space-time coordinates $x^\mu$ only, given by
\begin{equation}\label{eq:DeltaJ}
  \Delta_J = \left[\frac{\hbar}{a}\bm J - \frac{\kappa e a}{2c}\bm H\right]^2-
           \left[\frac{\hbar}{a}\bm K - \frac{\kappa e a}{2c}\bm E\right]^2.
\end{equation}
Here $\bm J$ and $\bm K$ are the generators of the Lorentz group in the dotted or undotted (conjugate) representation, according if $\psi^{\sigma'}_\sigma(x)$ or $\psi^{\dot\sigma'}_{\dot\sigma}(x)$ are considered. The connection with the spin $1/2$ Dirac's theory is made by taking $(u,v)=(0,\frac{1}{2})$ in Eq.~(\ref{eq:psiuv}) so that $D^{(0,1/2)}(\Lambda)\in SL(2,C)$. Then, introducing the Dirac four-spinor $\Psi_D=\pmatrix{\psi^{\sigma'}_{\sigma}\cr
\psi^{\dot\sigma'}_{\dot\sigma}}$ with $\sigma=\dot\sigma$ fixed and setting $\kappa=2$ for the electron, Eq.~(\ref{eq:coeff}) yields
\begin{widetext}
\begin{eqnarray}\label{eq:Dirac}
  \left[g^{\mu\nu}\left(\hat p_\mu -\frac{e}{c}A_\mu\right)
         \left(\hat p_\nu -\frac{e}{c}A_\nu\right)-
        \frac{e\hbar}{c}(\bm\Sigma{\ensuremath\cdot}\bm H -
        i\bm\alpha{\ensuremath\cdot}\bm E)\right]\Psi_D +  \nonumber\\
        \mbox{+} \left[\frac{e^2a^2}{c^2}(H^2-E^2)+
        \frac{3\hbar^2}{2a^2}(1+4\xi^2)\right]\Psi_D = 0,
\end{eqnarray}
\end{widetext}
where: $\bm\Sigma=\pmatrix{\bm\sigma & 0\cr 0 & \bm\sigma}$, $\bm\alpha=\pmatrix{\bm\sigma & 0\cr 0 & -\bm\sigma}$, and $\bm\sigma=\{\sigma_x,\sigma_y\sigma_z\}$ are the usual Pauli matrices. Setting $a=(\hbar/mc)\sqrt{3(1+4\xi^2)/2}$, where $m$ is the particle mass, and neglecting the term
$(ea/c)^2(H^2-E^2)=(ea/c)^2)(\frac{1}{2}F_{\mu\nu}F^{\mu\nu})$, Eq.~(\ref{eq:Dirac}) reduces to the square of Dirac's equation in its spinor representation, viz.~\cite[see, for example, Ref.~][Eq.~(32,7a)]{landau4}
\begin{equation}\label{eq:Diracgamma}
   \left[\gamma^\mu\gamma^\nu\left(\hat p_\mu -\frac{e}{c}A_\mu\right)
   \left(\hat p_\nu -\frac{e}{c}A_\nu\right)-m^2c^2\right]\Psi_D = 0,
\end{equation}
where $\gamma^\mu$ are Dirac's $4\times 4$ matrices in the spinorial representation. Equation (\ref{eq:Dirac}), comprehensive of the electromagnetic term proportional to $F_{\mu\nu}F^{\mu\nu}$,
was derived by Schulman by applying usual quantization rules to the relativistic top described by three Euler angles~\cite{schulman70}. In his work, Schulman proposed also generalized wave equations for fields
of arbitrary spin, which are equivalent to our Eqs.~(\ref{eq:coeff}) and (\ref{eq:DeltaJ}). We will refer to Schulman's paper for a detailed discussion about the physical implications of Eq.~(\ref{eq:Dirac}). However, it is worth noting that the term proportional to $F_{\mu\nu}F^{\mu\nu}$ in Eq.~(\ref{eq:Dirac}) can be cancelled out just transforming the Weyl curvature $R_W$ in the Lagrangian (\ref{eq:L0geo}) according to $R_W\rightarrow R_W-(ea/c\hbar\xi)^2(\frac{1}{2} F_{\mu\nu}F^{\mu\nu})$ so that Eq.~(\ref{eq:Dirac}) would reproduce the square of Dirac's equation exactly. Before concluding, we notice that the present approach applies to any spin (see Eq.~\ref{eq:psiuv}), as expected from a theory based on rotating fourlegs. When spin other than 1/2 are considered, the usual functional relationship for the particle mass $m(s)=a s(s+1)+b$, ($a,b$ constant) is found, because we considered a six internal degrees of freedom Lagrangian with no constraint. It would be then interesting to investigate if appropriate constraints can be imposed either to select the spin 1/2 only, as proposed in Ref.~\cite{balachandran80,balachandran}, or to extend the present fixed mass and spin approach to include a family of particles collected in a different Regge trajectory~\cite{chew61,atre86,biedenharn87}.\\
\section{Interpretation}
We derived the square of Dirac's spin 1/2 equation in the framework of the Affine Quantum Mechanics. The spinning particle was described as a conformal relativistic top, obtained from the minimal relativistic top introduced long time ago by Frenkel~\cite{frenkel26} and Thomas~\cite{thomas26} by formally replacing the mass with the Weyl curvature of the top configuration space. The dynamics of the conformal top is invariant under conformal changes of the metric. All trajectories of the conformal top are extremal curves of the corresponding Lagrangian and they can form bundles described by the Hamilton-Jacobi equation. In summary, according to the present AQM interpretation the quantum wave equation related to any quantum mechanical problem (here the Dirac's spin dynamics) should not be taken as the starting point of the theory, as usually done. It is rather a useful mathematical tool adopted to reduce a generally awkward nonlinear geometro-dynamical problem to a more tractable linear one, in order to obtain simultaneously the dynamical properties of the particle motion and the geometrical properties (Weyl's potential and curvature) that determine that motion. Then, according to AQM, the physical interpretation of the dynamical theory is different from the usual one since the particle trajectories here are the extremal curves of a Lagrangian, i.e. the geodesics of the Weyl's field acting on the particle~\cite{wheeler,*misner}. We may regard the space curvature as a ``guidance field'' which is ultimately imposed by the conformal invariance of the problem. A stochastic interpretation of the present theory is possible, as pointed out by~\cite{santamato84,*santamato85}. But it is not necessary since the particle motion on his trajectory can be traced with a precision allowed by the Heisenberg's uncertainty relations, which are of course valid in the AQM theory. We do believe that the present AQM geometro-dynamical interpretation of Quantum Mechanics leads to two consequences that may result of large relevance in modern Physics. First, several alleged ``mysteries of the quantum world'' may be partially unveiled or at least find an alternative plausible interpretation. Among them the quantum superposition and, hopefully, the entanglement and quantum nonlocality involved in the EPR paradigm. Moreover, the well known Penrose's claim for the role of gravitation in the state reduction process can be analyzed in concrete terms by the AQM theory by replacing the metric with the affine properties of space~\cite{penrose1,*penrose2}. Second, the AQM geometro-dynamical interpretation relates in an obvious and direct way the quantum world with general relativity, in particular with the modern paradigmatic endeavor of quantum gravity. Indeed it could suggest new research paths in that domain~\cite{penrose3}~\cite{wilczek05,*kiefer08,*wang06}~\cite{rovelli}.
\begin{acknowledgments}
We thank dott.\ Paolo Aniello for useful suggestions.
\end{acknowledgments}

\end{document}